\documentclass[12pt]{article}
\begin{document}

\begin{center}{\bf \Huge Feynman's sunshine numbers}\\[10pt]
{\Large David Broadhurst\footnote{Department of Physics
and Astronomy, The Open University, Milton Keynes MK7 6AA, UK}\\[10pt]
22 April 2010}\end{center}

{\bf Abstract}: This is an expansion of a talk for mathematics and physics
students of the Manchester Grammar and Manchester High Schools.
It deals with numbers such as the Riemann zeta value
$\zeta(3)=\sum_{n>0}1/{n^3}$.
Zeta values appear in the description of sunshine
and of relics from the Big Bang.
They also result from Feynman
diagrams, which occur in the quantum field theory of
fundamental particles such as photons, electrons and positrons.
My talk included 7 reasonably simple problems,
for which I here add solutions, with further details of their
context.

\section{Numbers}
\subsection{Sum of the first 136 cubes}

\begin{eqnarray*}
1^3&=&1\;=\;1^2\\
1^3+2^3&=&9\;=\;(1+2)^2\\
1^3+2^3+3^3&=&36\;=\;(1+2+3)^2\\
1^3+2^3+3^3+\ldots+136^3&\stackrel{?}{=}&
(68\times137)^2\;=\;(1+2+3+\ldots+136)^2
\end{eqnarray*}

It is rather easy to prove, by induction, that
\begin{equation}\sum_{n=1}^{N-1}n^3=\frac14N^2(N-1)^2
\label{E1}\end{equation}
for every integer $N\ge2$.

Assume that formula~(\ref{E1}) is true for one particular
integer, say $N=M$. Then
\[\sum_{n=1}^{M}n^3=\frac14M^2(M-1)^2+M^3=\frac14M^2(M+1)^2\]
and~(\ref{E1}) is true for $N=M+1$. It is true, by inspection,
for $N=2$, and hence for every integer $N>1$.

\subsubsection{Problem 1: proof by induction}

Prove that, for every integer $N>1$,
\begin{equation}\sum_{n=1}^{N-1}n^4 = \frac1{30}N(N-1)(2N-1)(3N^2-3N-1).
\label{E2}\end{equation}

\subsection{Sum of the first 136 inverse cubes}

\begin{eqnarray*}
\sum_{n=1}^2\frac1{n^3}&=&1+\frac18\;=\;
\frac{3^2}{2^3}\\
\sum_{n=1}^3\frac1{n^3}&=&1+\frac18+\frac1{27}\;=\;
\frac{251}{2^3\times3^3}\\
\sum_{n=1}^4\frac1{n^3}&=&
\frac{5\times11\times37}{2^6\times3^3}\\
\sum_{n=1}^6\frac1{n^3}&=&
\frac{7^2\times11\times53}{2^6\times3\times5^3}\\
\sum_{n=1}^8\frac1{n^3}&=&
\frac{31\times2538983}{2^9\times3\times5^3\times7^3}\\
\sum_{n=1}^{10}\frac1{n^3}&=&
\frac{11^2\times89\times359\times4957}{2^9\times3^6\times5^3\times7^3}\\
\sum_{n=1}^{12}\frac1{n^3}&=&
\frac{13^2\times151099201553}{2^9\times3^6\times5^3\times7^3\times11^3}
\end{eqnarray*}

It is quite hard to obtain the complete factorization of
the numerator of the sum of the first 136 inverse cubes. Let
\[S(N)=\sum_{n=1}^{N-1}\frac1{n^3}\]
then the numerator of $S(137)$ turns out to be
\[137^2\times359\times20273371\times20077753681\times95441340948666564707
\times P_{29}\times P_{47}\times P_{54}\]
where
\begin{eqnarray*}
P_{29}&=&11670683543184939914296499797\\
P_{47}&=&15515586948180802047607109239107700793654018233\\
P_{54}&=&135756874001680563462725618538328344771913356122240503
\end{eqnarray*}
are primes with 29, 47 and 54 decimal digits.

\subsubsection{Problem 2: modular arithmetic}

Find all the primes $p$ such that $p^2$ divides the numerator of $S(p)$,
using the result that for every prime $p$ and every
integer $n$ with $p>n>0$ there is a unique integer $m$
with $p>m>0$ and $m n-1$ divisible by $p$. [Hints follow.]

\subsubsection{Hints for Problem 2}

Let $p$ be any prime. Let $a$ and $b$ be
rational numbers with denominators that are not divisible by $p$.
If the numerator of $a-b$ is divisible by $p$,
we say that $a$ and $b$ are congruent modulo $p$.
Let's agree to abbreviate this by writing $a\equiv b\;({\rm mod}\;p)$.
Begin by showing that
\[2S(p)=\sum_{n=1}^{p-1}\left(\frac1{n^3}+\frac1{(p-n)^3}\right)
\equiv0\;({\rm mod}\;p).\]
Now show that $2S(p)/p\equiv-3T(p)\;({\rm mod}\;p)$, where
\[T(p) = \sum_{n=1}^{p-1}\frac1{n^4}.\]
We are told that for every integer $n$ with $p>n>0$
there is a unique integer $m$ with $p>m>0$ and
$m\equiv1/n\;({\rm mod}\;p)$. Hence you may now
use the result of Problem~1 to solve Problem~2.

\subsection{Riemann zeta values}

Now let's consider what happens as $N\to\infty$ in the sum
$S(N)=\sum_{N>n>0}1/n^3$. The answer has a name: $\zeta(3)$, where
$\zeta(s)$ is the value of the Riemann zeta function with argument $s$.
Bernhard Riemann (1826--1866) found remarkable relations between
the zeros of $\zeta(s)$ in the complex plane and the distribution
of prime numbers. Here, we shall study only the Riemann zeta values
\[\zeta(s)=\sum_{n=1}^\infty\frac1{n^s}\]
for integers $s>1$.

{\tt Pari-GP} is a computer algebra system designed for fast
computations in number theory and is available from
{\tt http://pari.math.u-bordeaux.fr}
for free. It works well on {\tt Windows} and {\tt linux} platforms.
Here, I ask {\tt Pari-GP} to print values of $\pi^s/\zeta(s)$ for
5 even integers $s\ge 2$, at 60-digit precision.

{\tt
default(realprecision, 60);\\
forstep(s=2,10,2, print([s, Pi$\hat{\phantom s}$s/zeta(s)]));
}

The output

{\tt
[2, 6.00000000000000000000000000000000000000000000000000000000000]\\{}
[4, 90.0000000000000000000000000000000000000000000000000000000000]\\{}
[6, 945.000000000000000000000000000000000000000000000000000000000]\\{}
[8, 9450.00000000000000000000000000000000000000000000000000000000]\\{}
[10, 93555.0000000000000000000000000000000000000000000000000000000]
}

strongly suggests (but does not prove) the exact evaluations
\begin{eqnarray*}
\zeta(2)&=&\frac{\pi^2}{6}\\
\zeta(4)&=&\frac{\pi^4}{90}\\
\zeta(6)&=&\frac{\pi^6}{945}\\
\zeta(8)&=&\frac{\pi^8}{9450}\\
\zeta(10)&=&\frac{\pi^{10}}{93555}
\end{eqnarray*}
which were proven in 1735 by the master analyst
Leonhard Euler (1707--1783).
For every {\em even} integer $s\ge2$, it is proven that
$\zeta(s)/\pi^s$ is a rational number.

For folk who love analysis (and only for them)
here is a cunning method to prove that $\zeta(2)=\pi^2/6$.
First one may use binomial
expansions to show that
\begin{eqnarray}
\zeta(2)&=&\int_0^1\int_0^1\frac{{\rm d}x\,{\rm d}y}{1-x y}
\label{E3}\\
\frac12\zeta(2)&=&\int_0^1\int_0^1\frac{{\rm d}x\,{\rm d}y}{1+x y}.
\label{E4}\end{eqnarray}
Adding these results and dividing by 2, we obtain
\begin{equation}
\frac34\zeta(2)=\int_0^1\int_0^1\frac{{\rm d}x\,{\rm d}y}{1-x^2y^2}.
\label{E5}\end{equation}
Then the delicate transformations
\begin{eqnarray}
x&=&\frac{\sin(a)}{\cos(b)}\label{E6}\\
y&=&\frac{\sin(b)}{\cos(a)}\label{E7}
\end{eqnarray}
give
\begin{equation}
\frac34\zeta(2)=
\int_0^\frac{\pi}{2}{\rm d}a\int_0^{\frac{\pi}{2}-a}
{\rm d}b\;=\;\frac{\pi^2}{8}
\label{E8}\end{equation}
and hence $\zeta(2)=\pi^2/6$.

\subsubsection{Problem 3: evaluating $\zeta(2)$}

Prove equations (\ref{E3}), (\ref{E4}), (\ref{E5}) and (\ref{E8}).
[This is for enthusiasts of analysis, for whom hints follow.
An apocryphal remark by Sherlock Holmes appears in the solution.]

\subsubsection{Hints for Problem 3}

Show that the integrals in (\ref{E3}) and (\ref{E4}) give
$1+\frac14+\frac19+\frac1{16}+\ldots$ and
$1-\frac14+\frac19-\frac1{16}+\ldots$.
Hence prove (\ref{E4}) by considering the difference of these sums.
Then work out the Jacobian of the transformations
(\ref{E6}) and (\ref{E7}),
namely the determinant
\begin{equation}J(x,y)=\left|\begin{array}{l r}
\partial x/\partial a&\partial x/\partial b\\
\partial y/\partial a&\partial y/\partial b\end{array}\right|,
\label{E9}\end{equation}
and obtain the result in (\ref{E8}) from~(\ref{E5}).

\subsection{Is $\zeta(3)/\pi^3$ an irrational number?}

It is proven that $\zeta(3)$ is not a rational number,
but it is not known, for certain, whether $\zeta(3)/\pi^3$
is irrational. It seems, however, overwhelmingly probable
that this number is irrational. There are rather good algorithms
with which to guess integers $N$ and $D$
such that $\zeta(s)/\pi^s$ is close to $N/D$.
For {\em odd} integers $s\ge3$, no such attempt has produced
a guess that works. For even integers $s\ge2$, this experimental
method works rather well.

\subsubsection{Problem 4: the PSLQ and LLL algorithms}

Use the {\tt lindep} procedure of {\tt Pari-GP}, at 100-digit
precision, to search for an integer relation between
the three constants in the vector
$[\zeta(12),\,\pi^{12},\,1]$. Check that the PSLQ and LLL
options of {\tt lindep} give the same result and that the
resulting formula for $\zeta(12)$ still holds at 1000-digit
precision. Finally, investigate, with comparable vigour,
whether there might be an integer relation between the
constants in the vector $[\zeta(13),\,\pi^{13},\,1]$. [This problem
is for enthusiasts of free-ware, for whom hints follow.]

\subsubsection{Hints for Problem 4}

In the problem, you are asked to perform
investigations for $\zeta(12)$ and $\zeta(13)$.
Here, I indicate how I used {\tt Pari-GP} to
investigate $\zeta(2)$ and $\zeta(3)$. I began with

{\tt default(realprecision, 100); flag = -3;\\
print(lindep([zeta(2), Pi$\hat{\phantom s}$2, 1],
flag)$\tilde{\phantom s}$);}

which gave the answer ${\tt [-6,\,1,\,0]}$.
This means that, at 100-digit precision,
the procedure {\tt lindep} ``guessed" the integer relation
$-6\times\zeta(2)+1\times\pi^2+0\times1=0$. The
flag $-3$ invokes the PSLQ algorithm of Helaman Ferguson and
David Bailey. Note that I introduced the ``red-herring" constant $1$
as a safety measure. Next, I changed the flag
to $0$, so as to invoke the LLL algorithm of
Arjen Lenstra, Hendrik Lenstra and L\'{a}szl\'{o} Lov\'{a}sz,
obtaining the same integer relation. To be even safer,
I evaluated $6\zeta(2)-\pi^2$ at 1000-digit precision,
obtaining the answer {\tt 0.E-1001}~. Then, I investigated
the vector $[\zeta(3),\,\pi^3,\,1]$, obtaining no credible
integer relation from either PSLQ or LLL.
The PSLQ algorithm sensibly refused to give me a relation,
returning instead a large real constant as an error message.
The LLL algorithm gave me a relation, but this was
soon shown to be untenable by increasing the precision.

\section{Sunshine (and Big Bang) numbers}

\subsection{The Stefan--Boltzmann constant}

A black body is one that absorbs electromagnetic
radiation with maximum efficiency
and emits it with maximum efficiency, at all frequencies $f$.
It appears black only when very cold. When it gets hot it will glow,
first red-hot and then at higher temperatures $T$ white-hot.
The spectrum of its radiation depends on $f$ only via the
convenient dimensionless ratio
\[x= \frac{h f}{k T}\]
where $h$ is Planck's constant and
$k$ is Boltzmann's constant.
If the temperature, $T$, of the black body (relative to absolute zero)
is given in Kelvins (K) and the frequency, f, is given
in Herz (Hz), then the approximate values $h=6.626\times10^{-34}$~J~s and
$k=1.381\times10^{-23}$~J~K$^{-1}$ will serve here.

The spectrum of sunlight that we receive at the top of the Earth's
atmosphere is rather similar to the spectrum produced by a black body
with $T\approx6000$~K. The much spikier
spectrum that we receive at sea level results from the
complex composition of our atmosphere and may be subject
to dramatic change, in your lifetime, by our carbon emissions.

Electromagnetic energy is emitted by a black body
of surface area $A$ and temperature $T$ at rate
given by $\sigma A T^4$, where
\begin{equation}
\sigma=\frac{2\pi^5}{15}\,\frac{k^4}{h^3c^2}
=5.670\times10^{-8}~{\rm J}~{\rm m}^{-2}~{\rm s}^{-1}~{\rm K}^{-4}
\label{E10}\end{equation}
is called the Stefan--Boltzmann constant. It has a
dependence on $h$, $k$ and the speed of light
$c=2.998\times10^8$~m~s$^{-1}$ that is easily found by
dimensional analysis. (Just look at the units of $\sigma$
and you can do it in your head.)

But where, on Earth, does that Stefan--Boltzmann factor
\[\frac{2\pi^5}{15}\approx40.80\]
come from, for goodness sakes?

Amusingly, it turns out that we don't need the precise value
of this constant to relate the temperature of the Earth, $T_{\rm E}$,
to the temperature of the surface of the Sun, $T_{\rm S}$. If we
assume that both the Earth and the Sun radiate like black bodies,
then equilibrium, on Earth, is achieved when
\[\sigma(4\pi R_{\rm E}^2)T_{\rm E}^4 \approx
\left(\frac{R_{\rm E}}{2D}\right)^2\sigma(4\pi R_{\rm S}^2)T_{\rm S}^4\]
where $R_{\rm E}$ and $R_{\rm S}$ are the radii of the Earth and Sun
and $D$ is the distance between them. Note that the first
factor on the right hand side is the fraction of the Sun's power
that reaches us and is determined simply by geometry. Thus
we don't need the values of $\sigma$ or $R_{\rm E}$
and arrive at the estimate
\[\frac{T_{\rm E}}{T_{\rm S}} \approx
\frac12\sqrt{\frac{2R_{\rm S}}{D}}
\approx\frac12\sqrt{\frac{32}{60}\frac{2\pi}{360}}\approx0.048\]
using only the average angular diameter of the Sun, which is about 32
minutes of arc. If we take $T_{\rm E}=288$~K (about 15 degrees Celsius)
as the average temperature of the Earth, then we estimate
that $T_{\rm S}\approx T_{\rm E}/0.048=6000$~K.

\subsection{Planck's integral}

In 1900, Max Planck (1858--1947) gave a formula
for the black-body spectrum, i.e.\ the probability
that the energy emitted by a black body lies
in the narrow range of frequencies between
$f$ and $f+{\rm d}f$. This is proportional to $B(x){\rm d}x$,
with $x=h f/(k T)$ and
\[B(x)=\frac{2\pi x^3}{\exp(x)-1}.\]

When I read about this, in 1962, it was clear that there
was an interesting integral to do. To derive the
factor of $40.80$ in the Stefan-Boltzmann constant,
one needs to prove that
\[\int_0^\infty\frac{2\pi x^3\,{\rm d}x}{\exp(x)-1}=\frac{2\pi^5}{15}\]
but the school library only seemed to contain
texts saying things like ``it can be shown that\ldots" this evaluation
is correct. There was no internet to consult.
Nor was there a convenient servant like this

{\tt default(realprecision, 60);\\
lhs = intnum(x=0,[[1],1], 2*Pi*x$\hat{\phantom s}$3/(exp(x)-1)); print(lhs);\\
40.8026246380375271016988413391247475307067608837433320737990\\
rhs = 2*Pi$\hat{\phantom s}$5/15; print(rhs);\\
40.8026246380375271016988413391247475307067608837433320737990}

available to a 15-year-old in the early 1960's,
to lend such numerical reassurance.

\subsubsection{Problem 5: the sunshine number $\zeta(4)$}

By expanding in powers of $\exp(-x)$ show that
\[\int_0^\infty\frac{2\pi x^3\,{\rm d}x}{\exp(x)-1}=12\pi\zeta(4).\]

\subsubsection{Hints for Problem 5}

You will need to prove that
\[\int_0^\infty x^3\exp(-n x)\,{\rm d}x=\frac{3!}{n^4}.\]
To do so, use the integration variable $y=n x$
and integrate by parts, three times.

\subsection{Evaluations of $\zeta(4)$}

Thus the problem of the power of sunshine has been pushed
back to Euler's 1735 result that $\zeta(4)=\pi^4/90$.

\subsubsection{$\zeta(4)$ from the cotangent function}

It is a deep result in complex analysis that we
have the wonderful formula
\begin{equation}
\frac{\cos(z)}{\sin(z)}=\sum_{n=-\infty}^{\infty}\frac{1}{z-n\pi}.
\label{E11}\end{equation}
The right hand side has the same singularities, at $z=n\pi$,
with the same unit residues as occur for the cotangent,
on the left hand side.
Thus the difference between the right and left hand sides
is an entire function, with no singularities in the complex plane.
That does not, of itself, mean that this difference is the
zero function, since $\exp(z)$ is an example of an
entire function that does not vanish.
Let's take~(\ref{E11}) on trust,
from deeper thinkers, and use it to show that $\zeta(4)=\pi^4/90$.
First we should multiply by $z$, to remove the singularity at $z=0$,
and then combine the terms with positive and negative $n$,
to obtain
\[\frac{z\cos(z)}{\sin(z)}=
1-2z^2\sum_{n=1}^\infty\frac{1}{n^2\pi^2-z^2}\]
with a sum that converges near $z=0$.
We know how to expand the left hand side, as $z\to0$,
using a ratio of very simple Taylor series for $\cos(z)$
and $\sin(z)$.
On the right hand side we find the constants
$\zeta(2k)/\pi^{2k}$, after binomial expansion.
In particular, by simplifying the relation
\[\frac{1-z^2/2!+z^4/4!+O(z^6)}{1-z^2/3!+z^4/5!+O(z^6)}=
1-2\,\zeta(2)\frac{z^2}{\pi^2}-2\,\zeta(4)\frac{z^4}{\pi^4}+O(z^6)\]
we easily prove Euler's results
that $\zeta(2)=\pi^2/6$ and $\zeta(4)=\pi^4/90$.
We may go further, if we have a convenient
servant to do the Taylor expansions for us:

{\tt print(Vec(1 - z*cos(z)/sin(z) + O(z$\hat{\phantom s}$14))/2);\\{}
[1/6, 0, 1/90, 0, 1/945, 0, 1/9450, 0, 1/93555, 0, 691/638512875, 0]}

shows the origin of Euler's prime $691$ in the numerator of
$\zeta(12)/\pi^{12}$.

\subsubsection{$\zeta(4)$ from Fourier analysis}

Here I shall be even sketchier, since the context is more
strongly related to my undergraduate studies than to my school work.
Joseph Fourier (1768--1830) provided a method for
representing a function, say $f(x)$, on a finite interval,
say $\pi\ge x\ge -\pi$,
by an infinite series of sinusoidal functions,
say $\cos(n x)$ and $\sin(n x)$, with coefficients
that may be evaluated from integrals
of the product of $f(x)$ and the sinusoidal functions.
His method looks much prettier if we use Euler's
famous formula
\[\exp({\rm i}z) = \cos(z)+{\rm i}\sin(z)\]
where ${\rm i}^2=-1$. But don't be put off by ``the square root of $-1$";
we shall not need it, to evaluate $\zeta(4)$.

There are not many questions on Fourier analysis
that can be solved straightforwardly, under examination
conditions. Normally, the student is asked to analyze a
very simple function, say $f(x)=x^2$,
for which the Fourier coefficients in the series
\begin{equation}
f(x)=\sum_{n=-\infty}^\infty a_n\exp({\rm i}n x)
\label{E12}\end{equation}
may be computed fairly easily. Now suppose that the student obtains
the correct Fourier coefficients, namely
\[a_n=\int_{-\pi}^\pi\frac{\exp(-{\rm i}n x)}{2\pi}f(x)\,{\rm d}x=
\int_0^\pi\frac{\cos(n x)}{\pi}x^2{\rm d}x=
\left\{\begin{array}{l r}
\pi^2/3&{\rm for}\;n=0\\
2(-1)^n/n^2&{\rm for}\;n\ne0\end{array}\right.\]
then s/he might be asked to substitute
$x=0$ in the Fourier series~(\ref{E12}) and obtain, thereby,
an evaluation of the Riemann eta value
$\eta(2)=\sum_{n>0}(-1)^{n-1}/n^2=\zeta(2)/2$.
This too is rather easy: at $x=0$ we obtain
$0=\pi^2/3-4\eta(2)$ from the Fourier series~(\ref{E12}),
using the result for the Fourier coefficients.

An interesting exam question might say:
``use Parseval's theorem to evaluate $\zeta(4)$".

The theorem in question gives an integral of the square of
any function $f(x)$ as the sum of the squares of its Fourier coefficients
$a_n$:
\[\frac{1}{2\pi}\int_{-\pi}^\pi|f(x)|^2{\rm d}x
=\sum_{n=-\infty}^{\infty}|a_n|^2.\]

\subsubsection{Problem 6: $\zeta(4)$ from Fourier and Parseval}

Use the given Fourier series for $x^2$ and Parseval's theorem
to prove that $\zeta(4)=\pi^4/90$.

\subsection{The sunshine number $\zeta(3)$}

We have seen that a black body of surface area $A$ and absolute
temperature $T$ radiates energy at a rate
\[\frac{{\rm energy}}{{\rm time}}=I_4 \frac{(k T)^4A}{h^3c^2}\]
where
\[I_4 = \int_0^\infty\frac{2\pi x^3\,{\rm d}x}{\exp(x)-1}
\;=\;12\pi\zeta(4)\;=\;\frac{2\pi^5}{15}\;\approx\;40.80\]
derives from an integral over the black-body spectrum.

Like all forms of electromagnetic radiation, light
is emitted and absorbed as ``bundles" of energy called photons.
The rate at which a black body of surface area $A$ and temperature $T$
emits photons is
\[\frac{{\rm photons}}{{\rm time}}=I_3 \frac{(k T)^3A}{h^3c^2}\]
where
\[I_3 = \int_0^\infty\frac{2\pi x^2\,{\rm d}x}{\exp(x)-1}
\;=\;4\pi\zeta(3)\;\approx\;15.11\]

\subsubsection{How many photons in your oven?}

\[\frac{{\rm photons}}{{\rm volume}}=
16\pi\zeta(3)\left(\frac{k T}{h c}\right)^3\]
where the constant is now $4I_3$ and the factor of $4$
takes account of the fact that only $\frac12$ of the
photons are moving towards the nearest wall and these have
an average value of $\frac12$ for the cosine of
the angle of incidence with which they hit the wall.

In the U.K.\ the formula for the absolute
temperature of an oven at ``gas mark"
$M\ge1$ is
\[T(M) = \left(273.15 + \frac{125}{9}(M-1)+135\right)~{\rm K}\]
since $M=1$ is defined as 135 degrees Celsius (identical to 275
degrees Fahrenheit)
and then each new mark increases the temperature by 25 degrees
Fahrenheit, i.e.\ by 125/9 degrees Celsius.
The zero of the Celsius scale is defined as $273.15$~K,
which is very close to the melting point of pure ice at standard
atmospheric pressure.

Let's set an empty 40-litre oven at gas mark 9 and assume
that the walls and door are perfectly black, on the inside.
After this oven has arrived at the design temperature, we
switch off the power. Working out
\[\frac{k T(9)}{h c} = \frac{1.381\times10^{-23}\times519.3}
{6.626\times10^{-34}\times2.998\times10^8}~{\rm m}^{-1}
=3.61\times10^4~{\rm m}^{-1}\]
we expect to have
\[16\pi\zeta(3)\times(3.61\times10^4)^3\times0.040 = 1.14\times10^{14}
\;{\rm photons}\]
inside the oven. When the oven eventually cools down to room temperature,
with $T\approx300$~K, the number will have
decreased by a factor of $(300/519)^3=0.193$, so now there will be
merely 22 trillion photons inside.

Now let's turn down the oven to $2.73$~K.

\subsection{How many photons in your Universe?}

About $13.7$ billion years ago, the whole of the Universe was
at a temperature comparable to the present temperature of the surface
of the Sun. Photons, electrons, protons (and other particles)
were in constant collision.
But then the expansion of the Universe (in fact
the expansion of space itself) cooled things down and when
the temperature fell below about $3000$~K the photons ceased to
interact significantly with matter. Since then, space has expanded
by a factor of about 1000, in each of its three dimensions,
and the cosmic background radiation that we readily detect nowadays
has a black-body spectrum with a temperature of about $3$~K.
The expansion of space has stretched out the most probable
wavelength, $\lambda=(2.898\times10^{-3}~{\rm m~K})/T$,
taking it from the visible region, at the temperature
of the surface of the Sun, to the microwave region, now.
In the last few years we have come to understand a great deal about
the hot Big Bang in which our Universe originated, by studying
this cosmic microwave background radiation. Its present temperature is
known rather precisely:
\[T_0=(2.728\pm0.002)~{\rm K}\]
where the subscript $0$ means ``now".

The whole of the Universe now has a background photon density
\begin{equation}
\frac{{\rm cosmic~photons}}{{\rm volume~of~Universe}}=
16\pi\zeta(3)\left(\frac{k T_0}{h c}\right)^3=4.12\times10^8~{\rm m}^{-3}.
\label{E13}\end{equation}
What ``volume" should we multiply by, to count the total number of photons?
For all we know, the Universe might have an infinite volume.
Yet we cannot see all of it. Let's consider just that
volume
\[V=\frac{4\pi}{3}R_0^3\]
 out to the present distance $R_0$
of the places from which we now receive light from the Big Bang.
Further than that we cannot see.
It is believed, nowadays, that we are allowed to use Euclidean geometry
in this calculation. Space-time is curved
but space seems not be, on a cosmic scale.

You might think that $R_0/c$ ought to be close to
$t_0=13.7\times10^9$~years,
the present age of the Universe.
In fact, it's not as simple as that: the curvature of space-time
in the early Universe means that we must use the general theory
of relativity. In the simplest current model,
\[\frac{R_0}{c t_0} =
\frac{I_1(\Omega_{\Lambda,0})}{I_2(\Omega_{\Lambda,0})}\]
is the ratio of two integrals that depend on
the crucial number $\Omega_{\Lambda,0}=0.734\pm0.02$, which is
the fraction of the present space-time curvature of the Universe
that comes from the cosmological constant.
(You may have heard this contribution referred
to, rather unhelpfully, as ``dark energy".) The integrals are
\begin{eqnarray*}
I_1(c)&=&\int_0^1\frac{{\rm d}a}{\sqrt{a^4c+(1-c)a}}\\
I_2(c)&=&\int_0^1\frac{a\,{\rm d}a}{\sqrt{a^4c+(1-c)a}}\;=\;
\frac{1}{3\sqrt{c}}\,\ln\left(\frac{1+\sqrt{c}}{1-\sqrt{c}}\right)
\end{eqnarray*}
and the first is too hard to do by hand. So let's ask our servant to do it:

{\tt default(realprecision, 5); c = 0.734; s = sqrt(c);\\
print(intnum(a=0,1, 1/sqrt(a$\hat{\phantom s}$4*c+(1-c)*a)));\\
3.4700\\
print(log((1+s)/(1-s))/3/s);\\
0.99677}

There are about are $3.156\times10^7$ seconds in a year.
So light travels $3.156\times10^7\times2.998\times10^8=9.46\times10^{15}$
metres in a year. Hence I claim that there are about
\[\frac{4\pi}{3}
\left(\frac{3.47}{0.997}\times13.7\times10^9\times9.46\times10^{15}\right)^3
\times4.12\times10^8=1.58\times10^{89}~{\rm photons}\]
present within the horizon beyond which you cannot possibly look.

Please remember that we needed the Riemann zeta value
\[\zeta(3)=\sum_{n=1}^\infty\frac{1}{n^3}=1.2020569\ldots\]
to do this cosmic calculation.

\subsection{How many neutrinos in your Universe?}

In nuclear beta-decay a neutron turns into a proton.
An electron and a anti-neutrino are ejected from the nucleus:
\[{\rm n}\longrightarrow{\rm p}+{\rm e}^-+\overline{\nu}_{\rm e}\]
sharing the released energy. We call $\overline{\nu}_{\rm e}$
the ``electron anti-neutrino". Other nuclei eject positrons
and electron neutrinos:
\[{\rm p}\longrightarrow{\rm n}+{\rm e}^++\nu_{\rm e}\]
A second pair of neutrinos is produced by the decays
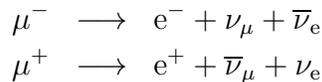
\begin{eqnarray*}
\mu^-&\longrightarrow&{\rm e}^-+\nu_\mu+\overline{\nu}_{\rm e}\\
\mu^+&\longrightarrow&{\rm e}^++\overline{\nu}_\mu+\nu_{\rm e}
\end{eqnarray*}
which occur in the Earth's atmosphere, where muons are
created by cosmic rays. A third pair, $\nu_\tau$ and $\overline{\nu}_\tau$,
is readily produce by the decays of ``tau-leptons", $\tau^\pm$,
at high energy physics laboratories. There seem to be no more.

There is a cosmic background
of neutrinos from the hot Big Bang with a ratio
\begin{equation}\frac{{\rm neutrinos}}{{\rm photons}}=3\,\frac{J_3}{I_3}\,
\left(\frac{T_\nu}{T_0}\right)^3\label{E14}\end{equation}
of the neutrino and photon densities, where $T_\nu\approx1.95$~K is the present
temperature of the neutrino background and
\begin{eqnarray*}I_3 &=& \int_0^\infty\frac{2\pi x^2\,{\rm d}x}{\exp(x)-1}\\
J_3 &=& \int_0^\infty\frac{2\pi x^2\,{\rm d}x}{\exp(x)+1}
\end{eqnarray*}
with a smaller integral $J_3$ for the neutrinos, because they
are particles that obey an ``exclusion principle".
Neutrinos, like electrons, utterly refuse to share quantum states with
particles identical to themselves. (This rather selfish behaviour
is fortunate for us, because it is the exclusion principle
for electrons that is responsible for the wonderful chemistry revealed
by the periodic table of the elements.)

The factor of 3 in the ratio~(\ref{E14}) comes from the three ``flavours"
of neutrino: e, $\mu$ and $\tau$. The next factor
comes from the exclusion principle.
In the simplest model, the final factor is a rational number:
$(T_{\nu}/{T_0})^3=\frac{4}{11}$.

\subsubsection{Problem 7: $\zeta(3)$ integral for cosmic neutrinos}

Show that $J_3=3\pi\zeta(3)$ and hence that there are, on
average, about 337 neutrinos, left over from the Big Bang,
in each cubic centimetre of your Universe. [Neutrinos
permeate matter with great ease. There may be more than 300
cosmic neutrinos in each cubic centimetre of your brain, as
you solve this problem, along with higher energy neutrinos
from local sources, such as the Sun and decays of muons in
the Earth's atmosphere.]

\section{Feynman's zeta values}

\subsection{Particles and fields}

On January 25, 1947, the journal Nature published a letter
from Donald Perkins~\cite{DHP} (n\'{e} 1925),
entitled ``Nuclear disintegration by meson capture".

On December 30, 1947, the Physical Review received a
one-page article by Julian Schwinger~\cite{JS} (1918--1994)
entitled ``On quantum-electrodynamics and the magnetic moment
of the electron".

Between January and December of that year came momentous
discoveries of two new types of particle, muons and pions,
and, quite independently, the stimulus for great leaps of
understanding in the quantum field theory of
electromagnetism, by Schwinger, Hans Bethe (1906--2005),
Richard Feynman (1918--1988) and Freeman Dyson (n\'e 1923).

\subsubsection{Particles}

Pions (or pi-mesons as they were called in 1947)
are created when high-energy protons from
distant regions of the Universe hit the Earth's atmosphere.
A negatively charged pion decays to give (most often)
a muon and a neutrino:
\[\pi^-\longrightarrow\mu^-+\overline{\nu}_{\mu}\]
The muon then decays to give an electron and a pair of neutrinos.
The discovery of pions and muons, in 1947, by the tracks that they
leave in photographic emulsion, was the beginning of the
experimental side of my subject: particle physics.
Since then many more types of particles have been found, in collisions
at particle accelerators. Currently, experimenters
at the large hadron collider in Geneva are searching for
a theoretically expected particle, called the ``Higgs boson".

\subsubsection{Fields}

James Clerk Maxwell (1831--1879) described the electromagnetic
fields, {\bf E} and {\bf B}, that are created by, and in turn influence,
charged particles. An example is the force between a pair of
current carrying wires, which is used to define the Amp\`ere (A),
as a unit of electrical current, in terms of the definition
of the Newton (N), as a unit of force. This definition appears
in Maxwell's equations via a constant
\[\mu_0\equiv\frac{4\pi}{10^7}~{\rm N}~{\rm A}^{-2}\]
which appears in magnetic-field calculations. Using this
value, we may define
a very important dimensionless number
\[\alpha\equiv\frac{e^2c}{2h}\mu_0 = \frac{1}{137.0359990\ldots}\]
called the fine structure constant. It is know to an accuracy
of about one part in a billion. Here $e^2$ is the square
of the electron's charge, $c$ is the speed of light and $h$
is Planck's constant. Physicists tend to remember the reciprocal
of the value of $\alpha$, which is close to $137$.

The quantum field theory developed in the late 1940's, by
Schwinger, Feynman and Dyson, combined
Maxwell's classical theory of electromagnetism (in which $\mu_0$
occurs) with Einstein's
special theory of relativity (in which $c$ is rather important)
and earlier quantum theory (in which $h$ is rather important).

Werner Heisenberg (1901--1976),
Erwin Schr\"{o}dinger (1887--1961)
and Paul Dirac (1902--1984) had developed the theory of quantum
mechanics, between 1925 and 1932. This confronts the fundamental
unpredictability of individual events, at the deepest level of
physics, and gives ways of calculating the probabilities of
outcomes, in a large number of measurements. But it does not handle
the creation and annihilation of particles. Schwinger and Feynman
solved that problem, in the interactions of photons, electrons and
positrons.

In the 1930's it was
realized that a ``photon field" was needed, to explain the creation
of photons by the interactions of charged particles, and that
an ``electron field" was needed, to explain the creation of
electrons and positrons by the photon field.

I shall try to indicate how the ``sunshine" number $\zeta(3)$,
and its cousins $\zeta(5)$, $\zeta(7)$, etc, appear in
the application of quantum field theory to particle physics.
But first, it would be a good idea to look at
the result of a rather simpler calculation.

\subsubsection{Electron-positron creation in pion decay}

The positron, e$^+$,
is the anti-particle of the electron, e$^-$,
with exactly the same mass, $m_{\rm e}$,
and exactly the opposite charge, $e$.

The neutral pion $\pi^0$ usually decays into a pair of photons:
\[\pi^0\longrightarrow\gamma+\gamma\]
where $\gamma$ is the symbol that we use for a photon (sometimes
called a gamma ray).
In the rest frame of the pion, each photon carries away an
energy $\frac12m_{\pi^0}c^2$, where $m_{\pi^0}$ is the mass of $\pi^0$.

In 1951, Richard Dalitz (1925--2006)
used the quantum field theory of Feynman and Schwinger
to predict another, rarer decay:
\[\pi^0\longrightarrow\gamma+{\rm e}^-+{\rm e}^+\]
in which an electron-positron pair is created, in place
of one of the photons. His predicted rate was a fraction~\cite{RHD}
\[{\rm Dalitz~pair~probability}\;=\;
\frac{\alpha}{\pi}\left(\frac43\ln\left(\frac{m_{\pi^0}}{m_{\rm e}}\right)
-\frac73\right)\;=\;1.185\%\]
of the total decays. This was a rather successful prediction.
The best result from modern experiments is that
$(1.198\pm0.032)\%$ of the decays of $\pi^0$ have the
electron-positron pair predicted by Dalitz.

Dalitz had to do an integral to make his prediction, since
the total energy released is here shared between 3
particles, in the final state, and we must integrate over
all the possible ways of doing that. In any individual
decay, we cannot say how the energy will be shared. We can
predict only the probabilities of the various outcomes and
check those predictions against a large number of
measurements.

\subsection{Magnetic moment of the electron}

\[\frac{\rm magnetic~moment}{\rm Bohr~magneton}=1+A_1\frac{\alpha}{\pi}
+A_2\left(\frac{\alpha}{\pi}\right)^2
+A_3\left(\frac{\alpha}{\pi}\right)^3+\ldots\]
where the Bohr magneton
\[\frac{e h}{4\pi m_{\rm e}} = 9.274\times10^{-24}~{\rm J}~{\rm T}^{-1}\]
has the units of Joules per Tesla and is the value predicted by Dirac
in 1928. In 1947, Schwinger found the first correction term, with~\cite{JS}
\[A_1=\frac12.\]
In 1950, Robert Karplus and Norman Kroll claimed the value~\cite{KK}
\[
28\zeta(3)
-54\zeta(2)\ln(2)
+\frac{125}{6}\zeta(2)
-\frac{2687}{288}
\;=\;
-2.972604271\ldots\]
for the coefficient of the next correction. It turned out
that they had made a mistake in this rather difficult calculation.
The correct result~\cite{SP1,SP2}
\[A_2\;=\;
\frac34\zeta(3)
-3\zeta(2)\ln(2)
+\frac{1}{2}\zeta(2)
+\frac{197}{144}
\;=\;-0.3284789655\ldots\]
was not obtained until 1959. Not until 1996 was the next coefficient
found~\cite{LR}
\begin{eqnarray}A_3&=&
-\frac{215}{24}\zeta(5)
+\frac{83}{12}\zeta(3)\zeta(2)
-\frac{13}{8}\zeta(4)
-\frac{50}{3}U_{3,1}\nonumber\\[3pt]&&{}
+\frac{139}{18}\zeta(3)
-\frac{596}{3}\zeta(2)\ln(2)
+\frac{17101}{135}\zeta(2)
+\frac{28259}{5184}\label{E15}\\[3pt]
&=&1.181241456\ldots\nonumber
\end{eqnarray}
by Stefano Laporta and Ettore Remiddi.
The irrational numbers appearing on the second line are those
already seen in $A_2$.
On the first line we see zeta values and
a new number, namely the alternating double sum
\[U_{3,1}=\sum_{m>n>0}\frac{(-1)^{m+n}}{m^3n}
\approx-0.1178759996505093268410139508341376187152\ldots\]

I visited Stefano and Ettore in
Bologna when they were working on this formidable
calculation and recommended to them a method of integration
by parts, in $D$ dimensions, that I had found useful for
related calculations in the quantum field theory of
electrons and photons. Here $D$ is eventually set to 4, the
number of dimensions of space-time. But it turns out to be
easier if we keep it as a variable until the final stage of
the calculation. Then if we find parts of the result with
factors like $1/(D-4)$, or $1/(D-4)^2$, we need not worry
that these parts tend to infinity as $D\to4$; all that
matters is that the complete result is finite. Based on my
$D$-dimensional experience~\cite{BQED}, I expected their final result to
look simplest when written in terms of the new number
$U_{3,1}$. This indeed turned out to be case. If, however,
one writes~(\ref{E15}) in terms of a less appropriate
non-zeta value, such as~\cite{BBBL}
\[\sum_{n=1}^\infty\frac{1}{n^4}\left(\frac12\right)^n
=-\frac{1}{24}[\ln(2)]^4 +\frac{1}{4}\zeta(2)[\ln(2)]^2
+\frac14\zeta(4)-\frac12U_{3,1},\]
then the additional irrational
numbers $[\ln(2)]^4$ and $\zeta(2)[\ln(2)]^2$ will also appear,
since they are, as I had expected, absent when one uses $U_{3,1}$.

\subsection{Feynman diagrams}

\vspace{-7mm}
\setlength{\unitlength}{0.0141cm}
\newbox\shell

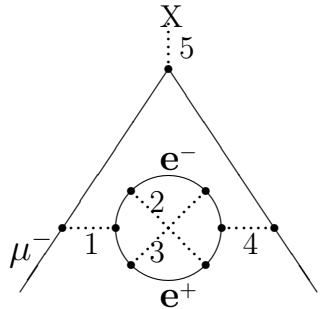
\def\pixel{\circle*{3}}
\def\vertx{\circle*{7}}
\def\gpict{\begin{picture}(400,390)(-100,-125)
           \put(0,0){\circle{100}}
           \put(-150,-30){{\large$\mu^-$}}
           \put(0,150){\line( 2,-3){140}}
           \put(0,150){\line(-2,-3){140}}
           \multiput(-50,0)(-7,0){8}\pixel
           \multiput( 50,0)( 7,0){8}\pixel
           \multiput(0,150)( 0,7){7}\pixel
           \put(   0,150){\vertx}
           \put( -50,  0){\vertx}
           \put(  50,  0){\vertx}
           \put( 100,  0){\vertx}
           \put(-100,  0){\vertx}}
\def\spict{\gpict
           \put(-80,-25){1}
           \put(-18, 15){2}
           \put(-18,-32){3}
           \put( 10,160){5}
           \put( -8,190){X}
           \put(+70,-25){4}
           \put( -8, 55){{{\large\bf e}$^-$}}
           \put( -8,-75){{{\large\bf e}$^+$}}}
\def\epict{\end{picture}}
\def\ntype{\multiput(0,0)( 4.9, 4.9){8}\pixel
           \multiput(0,0)( 4.9,-4.9){8}\pixel
           \multiput(0,0)(-4.9, 4.9){8}\pixel
           \multiput(0,0)(-4.9,-4.9){8}\pixel
           \put( 35, 35){\vertx}
           \put(-35, 35){\vertx}
           \put( 35,-35){\vertx}
           \put(-35,-35){\vertx}}
\def\align{\dimen0=\ht\shell
           \multiply\dimen0by7
   \divide\dimen0by16
   \raise-\dimen0\box\shell}

\hspace*{25ex}
\setbox\shell=\hbox{\spict\ntype
\epict}\align

Every Feynman diagram tells a possible story. In this one, a
muon emits photon 1, which creates an electron and a positron.
These exchange photons 2 and 3 and annihilate to give
photon 4. Meanwhile, the muon interacts with a
magnet X, via a (very low energy) photon 5, and then the
muon neatly absorbs photon 4, to emerge from this story
with (almost) the same energy and momentum as it started with.

This Feynman diagram specifies only one of the very large number
of integrals that one must do to compute the term of order $\alpha^4$
in the magnetic moment of the muon. Feynman found methods
that give this particular contribution as an 8-dimensional integral.

In 1995, Pavel Baikov and I found a method of obtaining a numerical accuracy
of 5 significant figures for such integrals~\cite{BB}. A previous method,
from Toichiro (Tom) Kinoshita~\cite{TK}, was shown to give errors of 5\%.

Tom is one of my true heros. He was born in 1925 and studied with
Sin-Itiro Tomonaga (1906--1979), who shared the 1965 Nobel Prize
in physics with Schwinger and Feynman.

\subsection{Multiple zeta values}

When there is a single very large external energy scale,
it is often a good approximation to neglect the masses of particles
inside a Feynman diagram. This occurs in the case of electron-positron
collisions at very high energy, which produce large numbers of
particles, such as protons and pions, containing light quarks.
So far all the results obtained from such diagrams have been
zeta-valued, with $\zeta(7)$ now appearing in recent results~\cite{BCK}.

I believe that this cannot continue, indefinitely. In 1995,
Dirk Kreimer and I calculated some diagrams in a ``toy"
theory that has only one type of interacting quantum field,
unlike the more pertinent electron-positron-photon and
quark-antiquark-gluon theories of the electromagnetic and strong
interactions. In this toy theory, we found Feynman diagrams
that give a {\em multiple} zeta value~\cite{BK,BGK,BBV}
\[\zeta(5,3)=\sum_{m>n>0} \frac{1}{m^5n^3}\]
which is the first double sum that cannot be reduced to
zeta values.

So I end with the conjecture that when practical quantum
field theory progresses just one more step in difficulty, to
calculations that yield the zeta value $\zeta(9)$, then those
same calculations will also yield the multiple zeta value $\zeta(5,3)$.

\section{Solutions}

\subsection{Solution 1}

Let $f(x)=\frac1{30}x(x-1)(2x-1)(3x^2-3x-1)$. Then simple algebra
shows that $f(x+1)-f(x)=x^4$. Hence it follows, by induction,
that $\sum_{N>n>0}n^4=f(N)$, for each integer $N\ge2$,
since this is true for $N=2$,

\subsection{Solution 2}

We replace $n$ by $p-n$ in the summation
$S(p)=\sum_{p>n>0}1/n^3$, add the result to $S(p)$ and divide by
the prime $p$
to obtain $2S(p)/p=\sum_{p>n>0}s(p,n)$, where
\[s(p,n) = \frac1p\left(\frac1{n^3}+\frac1{(p-n)^3}\right)=
\frac{p^2-3p n+3n^2}{n^3(p-n)^3}\equiv-\frac3{n^4}\;({\rm mod}\;p).\]
Hence $2S(p)\equiv0\;({\rm mod}\;p)$ and
$2S(p)/p\equiv-3\sum_{p>n>0}1/n^4\;({\rm mod}\;p)$.
For each integer $n$ with $p>n>0$ there is a unique integer $m$
with $p>m>0$ and $m\equiv1/n\;({\rm mod}\;p)$. Hence we replace
$n$ by $m$ and obtain $2S(p)/p\equiv-3\sum_{p>m>0}m^4\;({\rm mod}\;p)$.
Finally we use the result~(\ref{E2}) of Problem~1, which shows that
$30\sum_{p>m>0}m^4\equiv0\;({\rm mod}\;p)$ and hence that
$20S(p)/p\equiv0\;({\rm mod}\;p)$. Thus the numerator
of $S(p)$ is divisible by $p^2$ for every odd prime $p$
that is not equal to 5.

\subsection{Solution 3}

For $|z|<1$ we may use the binomial expansion
$1/(1-z)=\sum_{n>0}z^{n-1}$. With $z=x y$ under the integral
sign in~(\ref{E3}), this gives $\sum_{n>0} 1/n^2=\zeta(2)$, as claimed.
With $z=-x y$ in~(\ref{E4}), it gives
$\sum_{n>0}(-1)^{n-1}/n^2$. Call this latter sum $\eta(2)$
(a Riemann eta value). Then
$\zeta(2)-\eta(2)=2\sum_{k>0} 1/(2k)^2=\frac12\zeta(2)$
and hence $\eta(2)=\frac12\zeta(2)$, as claimed in~(\ref{E4}).
Now we compute the Jacobian in~(\ref{E9}), which gives
\begin{eqnarray*}J(x,y)&=&\frac{\cos(a)}{\cos(b)}\,\frac{\cos(b)}{\cos(a)}
-\frac{\sin(a)\sin(b)}{\cos^2(a)}\,
\frac{\sin(a)\sin(b)}{\cos^2(b)}\\
&=&1-\tan^2(a)\tan^2(b)=1-x^2y^2.\end{eqnarray*}
Hence we may replace
${\rm d}x\,{\rm d}y/(1-x^2y^2)$ in~(\ref{E5}) by
the infinitesimal product ${\rm d}a\,{\rm d}b$.
Finally we have to take care of the limits
for $a$ and $b$. We must impose the condition $a+b\le\frac{\pi}{2}$
to ensure that neither $x$ nor $y$ exceeds $1$.
Hence we obtain half of
the area of a square with side $\frac{\pi}{2}$, in~(\ref{E8}),
which proves that $\zeta(2)=\pi^2/6$.

[This is called an ``elementary" proof. As you have seen,
this does not mean that it is an ``easy" proof. Rather, it means
that the proof does not rely on some deep result in complex
analysis. Later, you will see proofs that may appear to be
much easier, yet rely on much deeper assumptions. When
Sherlock Holmes (in the movies, if not in Conan Doyle's books)
declares that his line of reasoning is
``elementary, my dear Watson", he does not mean that it is
trivial; rather he means that he can explain it, sometimes
at length, without relying on external authority. The same
type of thing occurs in mathematics. Frits Beukers, Eugenio Calabi
and Johan Kolk published this ``elementary" proof in 1993~\cite{BKC}.]

\subsection{Solution 4}

For $\zeta(12)$, both PSLQ and LLL should have given you Euler's result
\[\zeta(12)=\frac{691\pi^{12}}{638512875}\] with the
intriguing prime $691$ in the numerator. You should not,
according to current belief, have found any significant
integer relation between the constants
$[\zeta(13),\,\pi^{13},\,1]$. If you believe that you did,
then please check it at higher precision, where it will
almost certainly evaporate.

\subsection{Solution 5}

The black-body function may be written as
$B(x)=2\pi x^3\sum_{n>0}\exp(-n x)$. To evaluate
$I_4=\int_0^\infty B(x)\,{\rm d}x$ we make the substitution $x=y/n$,
in the $n$-th term, obtaining $I_4=2\pi\Gamma(4)\zeta(4)$,
with the $1/n^4$ term in $\zeta(4)=\sum_{n>0}1/n^4$ coming
from the change of variables and an overall factor of
\[\Gamma(z)=\int_0^\infty y^{z-1}\exp(-y)\,{\rm d}y\]
appearing at $z=4$. Integration by parts gives
$\Gamma(z)=(z-1)\Gamma(z-1)$. Moreover, $\Gamma(1)=1$.
Hence, for positive integer $n$, we obtain $\Gamma(n)=(n-1)!$
after $n-1$ integrations by parts. Thus $I_4=12\pi\zeta(4)$.

\subsection{Solution 6}
With $f(x)=x^2$, Parseval's theorem gives
$\pi^4/5 = (\pi^2/3)^2 + 8\zeta(4)$ and hence
$\zeta(4)=(1/5-1/9)\pi^4/8=\pi^4/90$.

\subsection{Solution 7}

The expansion of $1/(\exp(x)+1)=\sum_{n>0}(-1)^{n-1}\exp(-n x)$
gives a Riemann eta value in $J_3=4\pi\eta(3)$. Subtracting
$\eta(3)=\sum_{n>0}(-1)^{n-1}/n^3$ from
$\zeta(3)=\sum_{n>0}1/n^3$, we obtain
\[\zeta(3)-\eta(3)=2\sum_{k=1}^\infty\frac{1}{(2k)^3}=\frac14\zeta(3).\]
Thus $J_3/I_3=\eta(3)/\zeta(3)=\frac34$ and the neutrino to photon
ratio is $3\times\frac34\times\frac{4}{11}=\frac{9}{11}$.
We know from~(\ref{E12}) that there are about 412 photons
per cubic centimetre. Multiplying by $\frac9{11}$ we
obtain 337 neutrinos per c.c.

\section{Acknowledgments}

\subsection{The Manchester Grammar School: 1958--1964}

I learnt calculus from Eric Hodge (1904--1962) who sadly
died at the end of my fourth-form year. My enthusiasm for
the method of induction comes from him. The somewhat awesome
High Master of MGS at that time was Eric James (1909--1992)
who had taught Freeman Dyson at Winchester. Dyson's latest
book, {\em The Scientist as Rebel} (2006), is dedicated to
Eric and Cordelia James.

My attention to detail in mathematical calculations was
nurtured in the sixth form by Philip Scofield (1930--2010)
for whom a sharpened piece of chalk was the most precise
tool in all analysis. My physics teacher Aden Womersley
(1931--2007) had a lively interest in modern developments
and knew about quarks as soon as Gell--Mann hypothesized
them in 1964. Yet my most influential
teacher was John Scobell Armstrong (1927--2001) whose love
of enlightenment philosophy spoke to both my heart and my
head.

Around the time of the Cuban missile
crisis of late 1962, as I recall, I became fascinated
by the large factor $2\pi^5/15\approx40.80$ in the
Stefan--Boltzmann constant that determines how much sunshine hits
the upper atmosphere of planet Earth. Trying to understand
its origin seemed like a good diversion from wondering
whether Kennedy and Khrushchev were about to release
catastrophic amounts of nuclear energy.

\subsection{Zambia: 1965}

At the end of 1964, I left the Manchester Grammar School
and taught for 9 months at Munali Secondary School,
in Lusaka. The discipline of teaching physics and maths
to bright and highly motivated students did me a power of good.
One of my former students, the late Dr Wedson C.\ Mwambazi, became
the World Health Organization representative for Tanzania,
and later for Ethiopia.

\subsection{Oxford: 1965--1968}

I was very fortunate to have Heinrich Kuhn (1904--1994),
David Brink (n\'{e} 1930) and Patrick Sandars (n\'{e} 1935)
as my physics tutors at Balliol College. From Don Perkins'
lectures I discovered particle physics and became determined
to study this subject as a post-graduate student at the
newly established University of Sussex. I also heard in
Oxford how seriously Dick Dalitz regarded the quark model.

\subsection{Sussex: 1968--1971}

I was equally fortunate in my postgraduate teachers, Gabriel
Barton, David Bailin and Norman Dombey. It was fun, in my
first term, to work out how Schwinger had obtained the
correction term $\frac12(\alpha/\pi)$ for the magnetic
moment of the electron, back in 1947, the year of my birth.
Stan Brodsky visited Sussex and told me how amazingly
difficult it would be to obtain an exact value for the
coefficient of $(\alpha/\pi)^3$.

\subsection{Stanford: 1971--1973}

This good fortune continued during my post-doctoral years.
The Stanford Linear Accelerator Center (SLAC) in California
proved just the right place to be, when physicists
were discovering the ``parts" of the proton, by
scattering electrons off protons. Sid Drell and Jim
Bjorken were great enthusiasts of Feynman's ``parton model".
Ken Johnson (1931--1999) was visiting SLAC. He was already
sure that quantum field theory had come back to stay and would
soon make sense of the strong interactions
between pions and protons.

Yet it was also interesting to hear of competing metaphysics,
from Geoff Chew at Berkeley, across the Bay. Geoff thought that there
might be {\em no} quantum field theory of strong interactions.
Instead, he advocated a ``bootstrap" approach that might free
us from previous reliance on the ``undemocratic"
supposition of forces between a few types of fundamental particle,
somewhat in the manner of Baron von M\"unchhausen, who by legend
tried to defy gravity by pulling himself up by his own boot laces.

At Stanford, however, we desperately wanted Feynman's partons
to have the fractional charges and the ``three mathematical colours"
that had notionally been assigned to quarks. Those were
the glory days, when experiment was the ruler of our subject,
and we had to wait but a short while for extra
data, from electron-positron collisions and
neutrino-proton collisions, to confirm the quark-parton model
and open the flood gates to a new quantum field theory.

\subsection{Geneva: 1973--1974}

At the laboratory of the European Council for Nuclear Reseach,
CERN, in Geneva, I learnt from Gerard 't Hooft how quantum
field theory could also make sense of the weak interaction, responsible
for muon decay. Gerhard helped me to understand
why strong interactions get weaker and weak interactions get
stronger, at higher energies. Ken Johnson was right:
quantum field theory is definitely here to stay.

I came to appreciate just how much hard work there was to
do to meet the computational challenges of computing the
Feynman diagrams of these new theories. It was Gerhard's supervisor,
Martinus (Tiny) Veltman, who had begun to confront these challenges
in the early 1970's. In 1974 there appeared a review of Tiny's
computer-algebra package, {\tt Schoonship}, whose modern-day
successor {\tt Form} has been developed by Jos Vermaseren to be
a powerful engine which implements the subtle algorithms
that yield zeta-values like $\zeta(7)$ in current calculations
of massless Feynman diagrams.

\subsection{Oxford: 1974--1975}

Returning to Oxford as a junior research fellow, I revelled, with
Dick Dalitz, as wonderful new results from electron-positron annihilation
came from Stanford, and other laboratories, making the case for quarks
more and more compelling.

One day, when I was scheduled to give a colloquium on these
new results, a school-boy from Eton came to grill me. His
grasp of modern physics was remarkable and I wondered what
he might later contribute to my subject. In fact, after a
brief but distinguished period of work, Stephen Wolfram
forsook particle physics. His programme {\tt Mathematica}
has been marketed with great commercial vigour and is now
widely used by students and researchers to perform some of
the simpler computations in particle physics.

\subsection{The Open University: 1975--present}

When I joined the Open University, one of my first tasks was
to study Steven Weinberg's fine book, {\em Gravitation and
Cosmology}, and then help to write a course, {\em
Understanding Space and Time}. Julian Schwinger came to
Milton Keynes, to work with us on producing this course.
At my earnest request, he gave us a memorable 3-hour extempore
account of general relativity, as he perceived it. I could
scarcely believe my luck. I was getting paid to learn
things, for myself, and then was able to help to share those
good things with open-minded and enthusiastic students. My
own openness of mind is reinforced by my avocation, by my
partner Margaret and by our sons Stephen and Peter.

\subsection{Envoi}

To all those acknowledged above, and to many more, I give
humble and hearty thanks for all their goodness and loving
kindness.


\begin{thebibliography}{99}
\raggedright

\bibitem{BB}
P.\ A.\ Baikov and D.\ J.\ Broadhurst,
Proceedings of 4th International Workshop on Artificial Intelligence 
for High Energy and Nuclear Physics,
Pisa, Italy, 1995 (AIHENP, Pisa, 1995), 167--172.

\bibitem{BCK}
P.\ A.\ Baikov, K.\ G.\ Chetyrkin, J.\ H.\ K\"{u}hn,
Adler function, Bjorken sum rule, and the Crewther relation to order
$\alpha_s^4$ in a general gauge theory,
Phys.\ Rev.\ Lett., 104 (2010), 132004.

\bibitem{BKC}
F.\ Beukers, J.\ A.\ C.\ Kolk and E.\ Calabi,
Sums of generalized harmonic series and volumes,
Nieuw Archief voor Wiskunde, 11 (1993), 217-224.

\bibitem{BBV}
J.\ Bl\"{u}mlein, D.\ J.\ Broadhurst and J.\ A.\ M.\ Vermaseren,
The multiple zeta value data mine, 
Computer Physics Communications, 181 (2010), 582--625. 

\bibitem{BBBL}
J.\ M.\ Borwein, D.\ M.\ Bradley, D.\ J.\ Broadhurst and P.\ Lisonek,
Special values of multiple polylogarithms,
Transactions of the American Mathematical Society, 353 (2001), 907--941.

\bibitem{BQED}
D.\ J.\ Broadhurst,
Three-loop on-shell charge renormalization without integration:
$\Lambda _{\rm QED}^{\overline{\rm MS}}$ to four loops,
Zeit.\ Phys., C54 (1992), 599-606.
     
\bibitem{BGK}
D.\ J.\ Broadhurst, J.\ A.\ Gracey and D.\ Kreimer,
Beyond the triangle and uniqueness relations: non-zeta counterterms
at large $N$ from positive knots,
Zeit.\ Phys., C75 (1997), 559--574.

\bibitem{BK}
D.\ J.\ Broadhurst and D.\ Kreimer,
Knots and numbers in $\phi^4$ theory to 7 loops and beyond,
Int.\ J.\ Mod.\ Phys., C6 (1995), 519--524.

\bibitem{RHD}
R.\ H.\ Dalitz,
On an alternative decay process for the neutral $\pi$-meson,
Proc.\ Phys.\ Soc.\ London, A64 (1951), 667--669.

\bibitem{KK}
R.\ Karplus and N.\ M.\ Kroll,
Fourth-order corrections in quantum electrodynamics and the
magnetic moment of the electron,
Phys.\ Rev., 77 (1950), 536--549.

\bibitem{TK}
T.\ Kinoshita,
Improved evaluation of the $\alpha^3$ vacuum-polarization 
contribution to the $\alpha^4$ muon anomalous magnetic moment,
Phys.\ Rev., D47 (1993), 5013--5017.

\bibitem{LR}
S.\ Laporta and E.\ Remiddi,
The analytical value of the electron $(g - 2)$ at order $\alpha^3$ in QED,
Physics Letters, B379 (1996), 283--291.

\bibitem{DHP}
D.\ H.\ Perkins,
Nuclear disintegration by meson capture,
Nature, 159 (1947), 126--127.

\bibitem{SP2}
A.\ Petermann, Helv.\ Phys.\ Acta., 30 (1957), 407.

\bibitem{JS}
J.\ Schwinger,
On quantum-electrodynamics and the magnetic moment of the electron,
Phys.\ Rev., 73 (1948), 416.

\bibitem{SP1}
C.\ M.\ Sommerfield,
Magnetic dipole moment of the electron,
Phys.\ Rev., 107 (1957), 328--329.

\end{thebibliography}
\end{document}